\documentstyle[graphicx,pifont,amssymb,amsmath]{mn}

\title[IR Study of NGC\,3256]
{An infrared study of the double nucleus in NGC\,3256}

\author[Lira etal.]
{P.~Lira,$^{1}$ G.~Valentino,$^{1,2}$ M.~Ward,$^3$ S.~Hoyer,$^{1}$\\
$^1$ Departamento de Astronom\'{\i}a, Universidad de Chile, Casilla 36-D, Santiago, Chile\\
$^2$ Department of Astronomy \& Astrophysics, UC Santa Cruz, 201 Interdisciplinary Sciences Building,  Santa Cruz, CA 95064, USA\\
$^3$ Department of Physics, University of Durham,  Science Laboratories,  South Road,  Durham DH1 3LE, UK\\}

\begin{document}

\maketitle

\begin{abstract} 

We present new resolved near and mid-IR imaging and N-band
spectroscopy of the two nuclei in the merger system NGC\,3256, the
most IR luminous galaxy in the nearby universe. The results from the
SED fit to the data are consistent with previous estimates of the
amount of obscuration towards the nuclei and the nuclear star
formation rates.  However, we also find substantial differences in the
infrared emission from the two nuclei which cannot be explained by
obscuration alone. We conclude that the northern nucleus requires an
additional component of warm dust in order to explain its
properties. This suggests that local starforming conditions can vary
significantly within the environment of a single system.

\end{abstract}

\begin{keywords}
galaxies: general -- galaxies: nuclei
\end{keywords}

\section{Introduction}

The dominance of obscured star formation in the early Universe makes
the dusty nearby systems interesting laboratories in which to probe
their physics and evolution. Because of their high level of
obscuration, such studies are well suited to observations at near and
mid-IR wavelengths. Our understanding of these sources has improved
considerably since the IRAS era of the 1980's. This is partly due to
the results from ISO in the 1990's and most recently from Spitzer.
However, high spatial resolution is still not atainable from space
facilities. This situation is, however, rapidly changing with the
advent of new ground based instrumentation which is a valuable
complement to the space missions.

Because of its closeness ($D \sim 39$ Mpc, for H$_{\circ} = 70$ km
s$^{-1}$ Mpc$^{-1}$) and luminosity ($L_{bol} \sim L_{IR} \sim 6
\times 10^{11} L_{\odot}$), NGC\,3256 is one of the best studied
Luminous Infrared Galaxies (LIRGs), note: it is almost a ULIRG!  Its
powerful infrared emission is the result of a merger of two massive
galaxies, whose nuclei have a projected separation of only $\sim 1$
kpc, indicating an advanced stage of the collision.

NGC\,3256 is also amongst the most X-ray luminous galaxies for which
there is no clear evidence of the presence of a (luminous) active
nucleus. In fact, the estimated luminosity of NGC\,3256 ($L_{\rm x}
\sim 10^{42}$ ergs s$^{-1}$ in the 0.5-10 keV band - Lira et al.,
2002; Moran, Lehnert \& Helfand, 1999) is consistent with the maximum
limit normally adopted to separate AGN from starburst galaxies found
in deep X-ray surveys (see Alexander et al. 2005, and references
therein.)

In this paper we present a study of the near and mid-IR resolved
emission from the two nuclei in the prototype LIRG NGC\,3256. In
section 2 we describe the high-spatial resolution, ground-based
observations obtained in the N-band. In section 3 and 4 the details of
our data analysis is presented. In Section 5 the SED of NGC\,3256 is
explored using our new N-band data, plus observations extracted from
the literature and Spitzer and HST archival images. Finally, the
discussion of our results is presented in Section 6.

\section{Observations and data reduction}

\subsection{Imaging}

\setlength{\unitlength}{1cm}
\begin{figure}
\centering
\begin{picture}(20,12){
\put(-1,0){\includegraphics[scale=0.55]{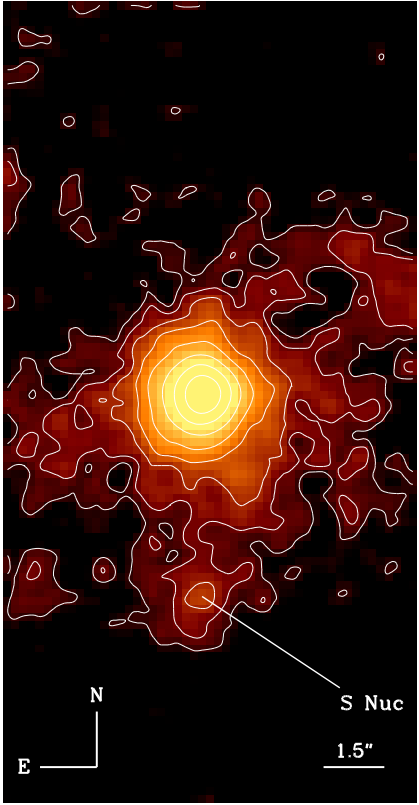}}}
\end{picture}
\caption{The high-resolution narrow N-band image ($\lambda_{c} \sim 11.5
\mu$m) of the central region of NGC\,3256 observed using 
TIMMI2. This image has been smoothed and displayed on a logarithmic
scale. Clearly the emission is dominated by the northern nucleus, but
the southern nucleus is also detected.}
\end{figure}

\begin{figure}
\centering
\includegraphics[angle=270,scale=0.45]{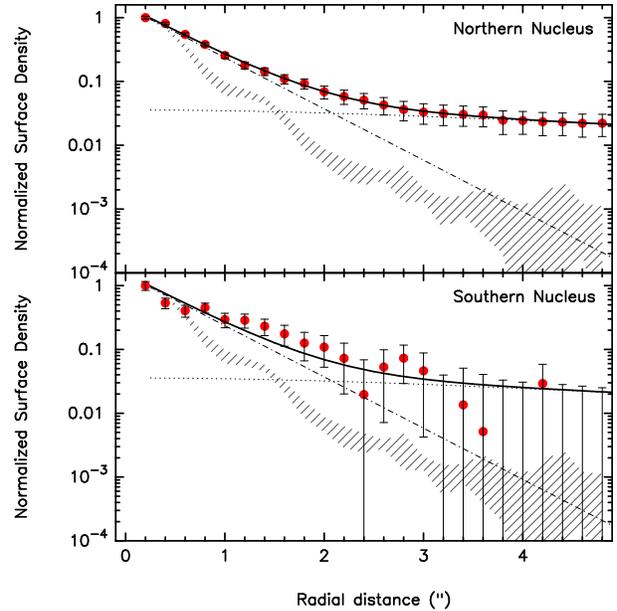}
\caption{Normalized radial profiles of the two nuclei in
NGC\,3256. The profile of the southern nucleus was derived by using
the lower half of the circular apertures centred on the nucleus, in
order to avoid the strong contamination from the northern source. A
1$\sigma$ interval in radial profiles of the standard stars observed
during the same night are also included for comparison (hashed
region). The extended profile of the northern nucleus was fitted using
the sum of a lorentzian and exponential function (shown as a thin
continuous line).  The same fit is superimposed on the profile of the
southern nucleus.}
\end{figure}

The high-resolution N-band imaging of NGC\,3256 was obtained with the
Thermal Infrared MultiMode Instrument (TIMMI2) camera on the ESO 3.6m
telescope at La Silla on the 24th of February 2001. A narrow N-band
filter was used ($\lambda_{c} \sim 11.5 \mu$m; $\Delta \lambda \sim
1.2 \mu$m), and the image scale was 0.2\arcsec\ per pixel. Standard
chopping and nodding techniques were employed with perpendicular chop
and nod throws of 15\arcsec. Despite the highly peaked light
distribution centered on the nuclear region, it is expected that some
galactic emission is also present $\sim 15\arcsec$ away from the
nuclear sources. This might affect our ability to obtain an accurate
sky subtraction, and is discussed further in Section 3.

The final image was constructed by shifting each individual frame onto
a common reference position determined by using the centroid of the
central source. The resulting exposure time was 1.16 ksecs. This image
was flux calibrated using standard stars observed throughout the
night. These showed an RMS error of 8\% in their fluxes. Inspection of
the radial profile of the standard stars showed that the observations
were diffraction-limited with a FWHM $\sim 0.8\arcsec$.

An uncalibrated, N-band 10.5 $\mu$m image of NGC\,3256 was published
by B\"oeker et al.~(1997) as one of the first science images obtained
with the MANIAC mid-IR camera. Siebenmorgen, Kr\"ugel \& Spoon (2004)
presented 8.6 $\mu$m and 10.4 $\mu$m imaging of this galaxy obtained
with TIMMI2, but the detection of the southern nucleus is not
discussed. Alonso-Herrero et al.~(2006) have recently published a
narrow-band, 8.7 $\mu$m image of NGC\,3256, and their measurements of
the northern and southern nuclei will be used later in Section 5. We
have also analysed archival Spitzer images of NGC\,3256 (PI.,
Fazio). These observations were carried out with the IRAC camera in
June 2004. The 'Post Basic Calibrated Data' were used to determine the
photometry of the nuclei. Saturation affected the northern nucleus in
the $8\mu$m band images, and so these were not used. Finally, for the
northern nucleus we also extracted the archival HST observations
obtained using the WFPC2 camera with the F300W, F555W and F814W
filters during June and Aug 2001 (PI's., Windhorst and van der
Marel). We did not include the optical data published by L\'{\i}pari
et al.~(2000) in our analysis because the used apertures and seeing
conditions did not meet our requirements (see Section 5).

\begin{figure*}
\centering
\includegraphics[scale=0.3]{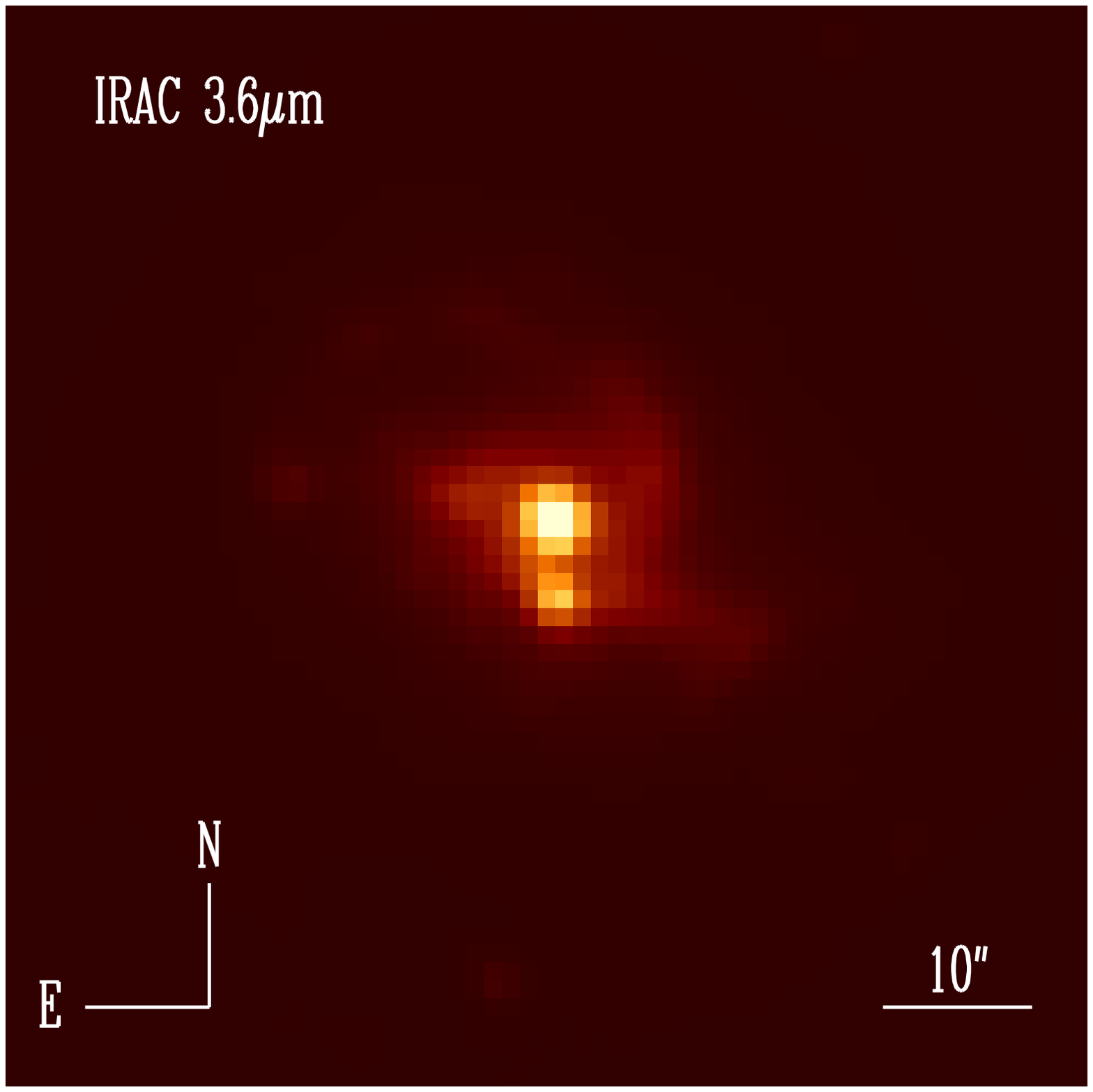}%
\includegraphics[scale=0.3]{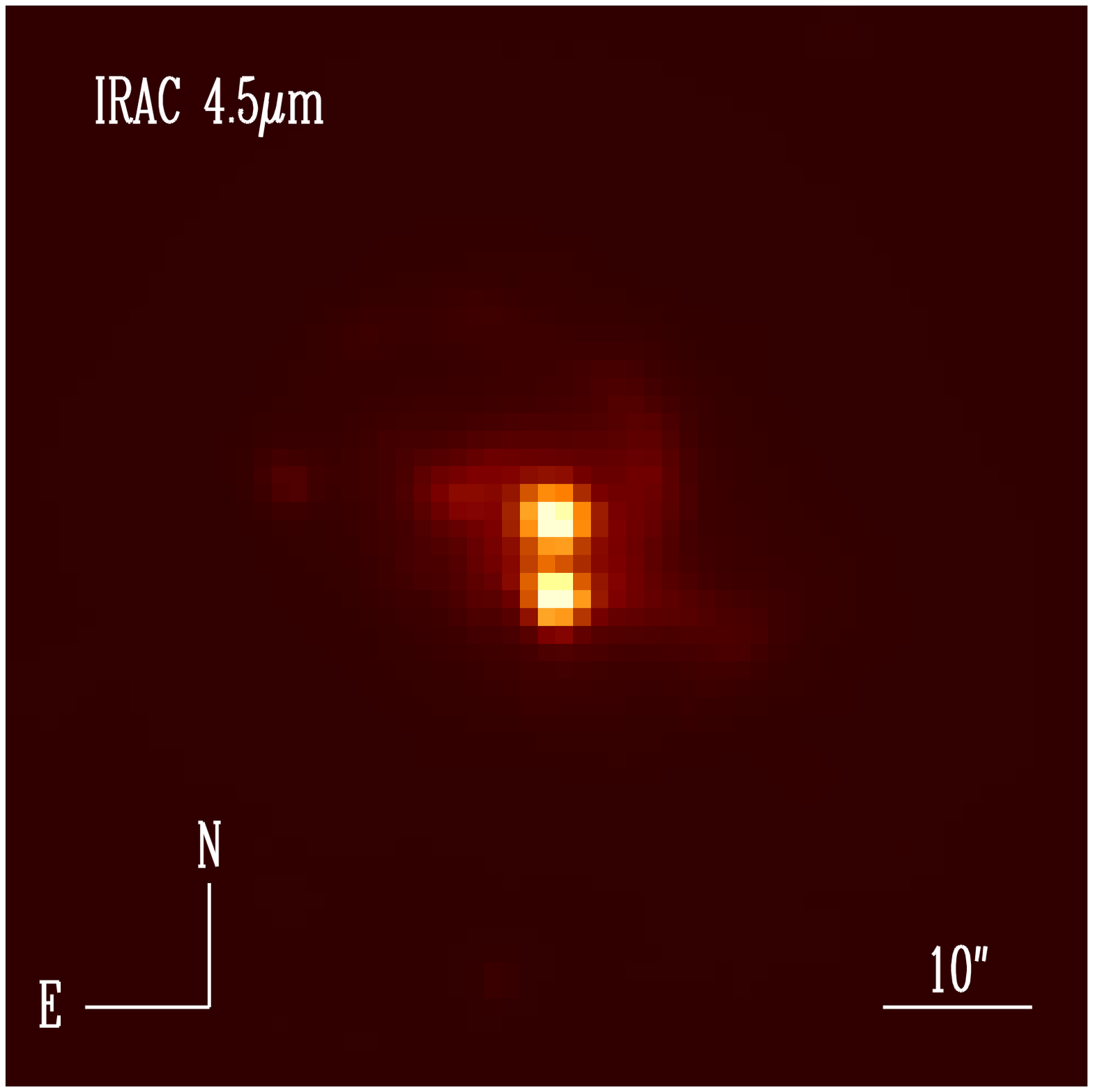}%
\includegraphics[scale=0.3]{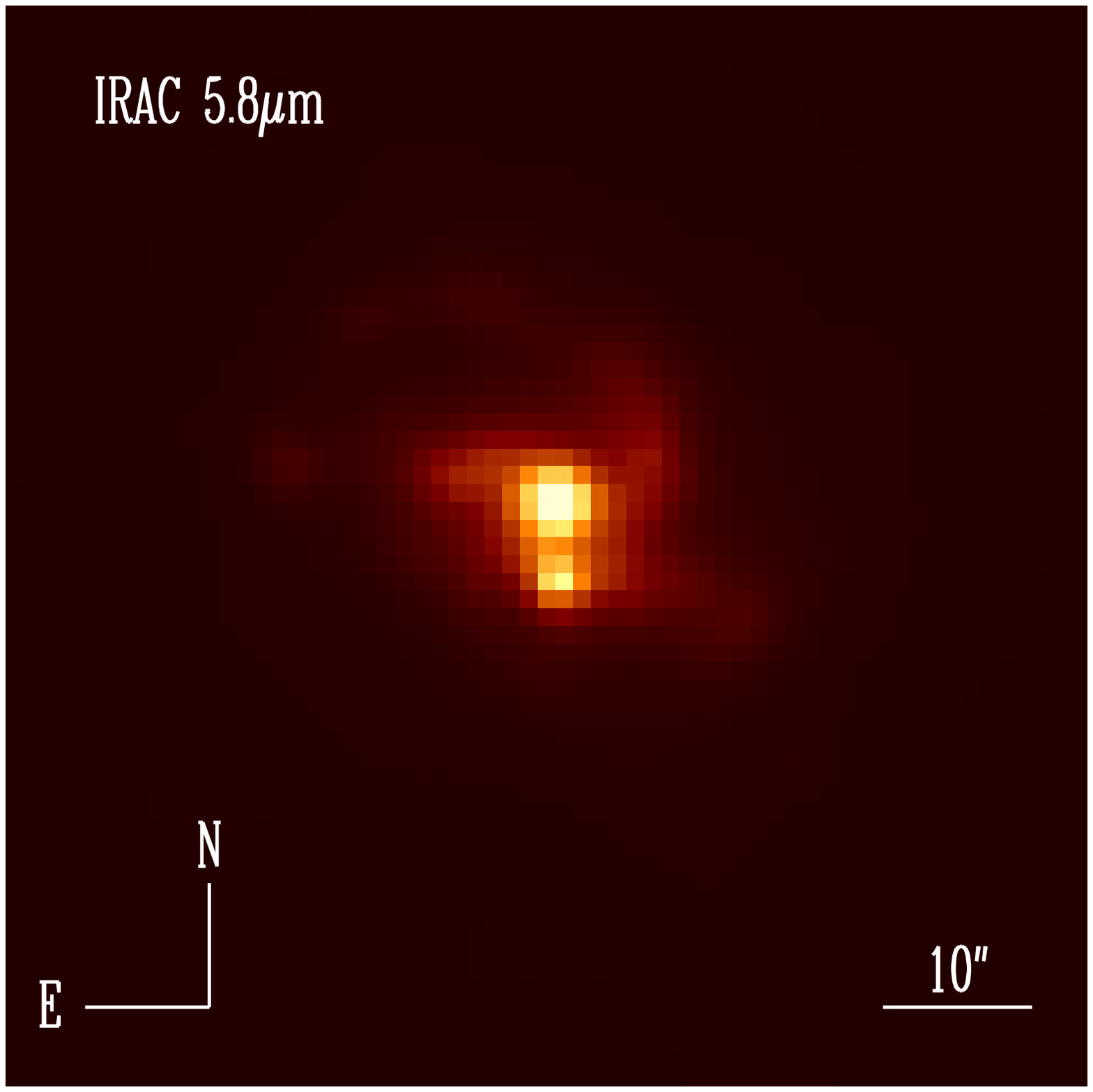}
\caption{IRAC images of NGC\,3256 in the 3.6, 4.5 and 5.8 $\mu$m bands.}
\end{figure*}

\subsection{Spectroscopy}

Long-slit N-band spectroscopy of NGC\,3256 was obtained during the
science verification phase of the TReCS camera on Gemini South on the
3rd of January 2004.  The low resolution KBrC grating was used giving
a spectral resolution $R \sim 100$ in the $8-12.5 \mu$m wavelength
range. The slit width was 1.3\arcsec\ and was rotated to a nearly N-S
position angle in order to cover both nuclei simultaneously.  The
chopping and nodding were performed in the direction perpendicular to
the slit orientation, with a throw of 30\arcsec. The total on-source
integration time was 1.28 ksecs.

The images were reduced using the Gemini {\it midir\/} IRAF package,
which derives background subtracted images from different chopping and
nodding positions, then registers them and produces a final averaged
image. Extraction of the spectra of both nuclei was performed using
standard IRAF package routines. Due to the low S/N available for the
southern nucleus, we used the trace function determined for the
northern nucleus to extract this spectrum. Aperture photometry of the
TIMMI2 image within a diameter of 1.3\arcsec\ has been used to
determine a good absolute flux calibration for the spectra.

Spitzer-IRS observations of NGC\,3256 have been published by Brandl et
al.~(2006) and are included in our analysis (Section 5). The
observations were carried out on the 13th of May 2004 using the
low-resolution module ($R \sim 65-130$). The slit widths were $\sim
3.6\arcsec$ in the $5-15\mu$m range, and $\sim 10.5\arcsec$ for
wavelengths grater than 15$\mu$m, although the final flux calibration
was obtained through the widest slit. Also, Mart\'{\i}n-Hern\'andez et
al.~(2006) have recently presented spatially resolved $7.5-13.9\mu$m
spectroscopy of the double nuclei in NGC\,3256 obtained with
TIMMI2. Finally, Siebenmorgen, Kr\"ugel \& Spoon (2004) presented
TIMMI2 and ISO spectroscopy for this galaxy, but as was the case with
their TIMMI2 imaging, there is no discussion of the possible detection
of the southern nucleus.

\section{Image analysis}

\begin{table*}
\centering
\caption{Aperture photometry of the two nuclei in NGC\,3256. Errors
correspond to 2\%\ for the WFPC2, 15\%\ for the IRAC, and 8\%\ for the
TIMMI2 measurements. Kotilainen et al.\ (1996) report errors of 5\%\ for
JHK' and 10\%\ for the L'-band photometry, respectively. See text
for details of the used apertures.}
\begin{tabular}{rcccl} \hline
Wave. & Ap.\ (\arcsec) & North (mJy) & South (mJy) & Source/Reference\\ \hline
2924 \AA\   & 3.0          & 0.45 $\pm$ 0.01 & --             & WFPC2 F300W - this work \\
5252 \AA\   & 3.0          & 2.79 $\pm$ 0.06 & --             & WFPC2 F555W - this work  \\
8269 \AA\   & 3.0          & 7.02 $\pm$ 0.14 & --             & WFPC2 F814W - this work  \\
1.25 $\mu$m & 3.0          & 21.6 $\pm$ 1.1  & 4.7 $\pm$ 0.2  & Kotilainen et al.\ (1996) \\
1.65 $\mu$m & 3.0          & 33.2 $\pm$ 1.7  & 9.0 $\pm$ 0.4  & Kotilainen et al.\ (1996) \\
2.10 $\mu$m & 3.0          & 33.7 $\pm$ 1.7  & 10.5 $\pm$ 0.5 & Kotilainen et al.\ (1996) \\
3.6  $\mu$m & 3.6          & 33.5 $\pm$ 5.0  & 16.8 $\pm$ 2.5 & IRAC band-1  - this work \\
3.75 $\mu$m & 3.0          & 32.3 $\pm$ 3.2  & 24.8 $\pm$ 2.5 & Kotilainen et al.\ (1996) \\
4.5 $\mu$m  & 3.6          & 30.0 $\pm$ 4.5  & 26.0 $\pm$ 3.9 & IRAC band-2  - this work \\
5.8 $\mu$m  & 4.0          & 128  $\pm$ 19   & 71 $\pm$ 11    & IRAC band-3  - this work \\
8.74 $\mu$m & 1.4$^{\star}$& 230  $\pm$ 11   & 40 $\pm$ 3     & Alonso-Herrero et al.\ (2006) \\
11.5 $\mu$m & 1.3$^{\dagger}$& 350 $\pm$ 28  & 26 $\pm$ 2    & TIMMI2  - this work \\ 
11.5 $\mu$m & 3.0          & 810  $\pm$ 65   & 86 $\pm$ 7    & TIMMI2  - this work \\ 
11.5 $\mu$m & 9.5          & 1800 $\pm$ 144  & --            & TIMMI2  - this work \\ \hline
\multicolumn{5}{l}{$^{\star}$ A factor 1.4 is required to correct these fluxes to a $\sim 3.0\arcsec$ aperture.}\\
\multicolumn{5}{l}{$^{\dagger}$ Used to derive the absolute calibration for the TReCS spectroscopy (see Fig.~4).}\\

\end{tabular}
\end{table*}

Figure 1 shows the TIMMI2 observations of the nuclear region in
NGC\,3256. The emission is clearly dominated by a central extended
source, although other clumps of emission can be seen distributed
around it. We identify the central source with the northern nucleus,
while the much fainter source located $\sim 5\arcsec$ to the south,
corresponds to the southern nucleus.

The radial profiles of both nuclei are shown in Figure 2. The northern
nucleus is clearly extended when compared with a stellar PSF. A best
fit to its profile was obtained by using a combination of an
exponential and a lorentzian function. In order to reduce
contamination from the northern nucleus in the photometry of the
southern source, its profile was derived using photometry obtained
from the lower-half of the circular apertures centred on this
source. Despite the much lower signal-to-noise, it is clearly seen
that the southern nucleus is also well resolved. Furthermore, the same
fit that described the profile of the northern nucleus is also found
to provide a good representation of the southern source. In fact the
HST-NICMOS images show that the southern nucleus is the more extended
of the two nuclei at near-IR wavelengths (Lira et al., 2002).

Photometry calculated within an aperture of 3\arcsec\ diameter centred
on the northern and southern nuclei, gives fluxes of $\sim 0.8$ and
$\sim 0.1$ Jy, respectively (see Table 1). Despite our efforts to
reduce the contamination from the northern nucleus in the southern
aperture, we must consider the flux from the southern nucleus to be
uncertain. The total emission within an aperture of 9.5\arcsec\ in
diameter centred on the northern peak corresponds to 1.8 Jy. However,
since the background subtraction of the image was obtained using a
chopping throw of only 15\arcsec, it is important to quantify the
possibility of sky oversubtraction which could affect these
measurements.

The total emission measured from the TIMMI2 image can be compared with
the IRAS 12$\mu$m ($\Delta \lambda = 8.5-15 \mu$m) photometry obtained
by Sanders et al. (1995). For an IRAS beam of 0.77\arcmin\ in diameter
they measured a marginally resolved source with a total flux of 3.6
Jy, which corresponds to a corrected flux of $\sim 5$ Jy in the
$10.9-12.1 \mu$m TIMMI2 band, if the northern nucleus best-fit model
presented in Section 5 is adopted, and this model is representative of
the emission from the whole region. Hence, the IRAS observations
provide an estimate of the size of the total emitting region, which is
comparable to the IRAS beam.

If we assume that the differences between the IRAS and the TIMMI2
fluxes are due to emission evenly distributed within an annular region
of inner and outer diameters 9.5\arcsec and 0.77\arcmin, respectively,
then the corresponding surface brightness would be $\sim 2$
mJy/arcsec$^{2}$. This would correspond to corrections of $\sim 2\%$,
$\sim 15\%$, and $\sim 10\%$ respectively to the fluxes for the total,
northern and southern regions presented in Table 1 (see 3 last
entries). However, it is quite possible that the emission from outside
the TIMMI2 image does not follow a completely even distribution, but
instead has a steep profile. To test this scenario we extrapolated the
fitted profile shown in Figure 2 out to the region used for the sky
subtraction ($\sim 15\arcsec$ away). The estimated value is only $\sim
0.3$ mJy/arcsec$^{2}$. Again, this result suggests that our
measurements do not suffer from a severe background
oversubtraction. Integrating the fit to the profile of the northern
nucleus out to a diameter of 0.77\arcmin\ underestimates the observed
IRAS flux by $\sim 60\%$, suggesting that the larger scale surface
brightness (i.e., that lying outside the TIMMI2 image) is actually
closer to the flat distribution that we proposed earlier.

Figure 3 presents the IRAC images in the 3.6, 4.5 and 5.8 $\mu$m
bands. The spatial resolution of IRAC (FWHM $\sim 1.5\arcsec$) clearly
separates the two nuclei. It can be seen that in the 4.5 $\mu$m band
image the two nuclei have a very similar peak brightness, while the
northern nucleus is more prominent at 3.6 and 5.8 $\mu$m. 

Ptak et al.~(2006) and Taylor-Mager et al.~(2007) have recently
presented the archival WFPC2 data used in this work to obtain the
optical photometry for the northern nucleus. Also, a spectacular WFPC2
colour image of NGC\,3256 can be found in Zepf et al.~(1999).

The TIMMI2, IRAC and WFPC2 photometry is given in Table 1. The IRAC
photometry was obtained in apertures that roughly correspond to the
same spatial region as measured in the 3.0\arcsec\ TIMMI2 aperture
photometry. This was achieved by deconvolving the TIMMI2 image of the
northern nucleus (using a standard star observation) and then
convolving it with the characteristic FWHM for each IRAC
band. Finally, an aperture size was defined so that the same
fractional flux as that measured from the TIMMI2 image would be
obtained. We assumed that the same IRAC apertures apply for the
photometry of the southern nucleus. Given the small FWHM of the WFPC2
camera ($\sim 0.15\arcsec$) a 3.0\arcsec\ aperture was directly
applied to obtain the optical photometry for the northern nucleus.

\section{Spectral analysis}

\begin{figure}
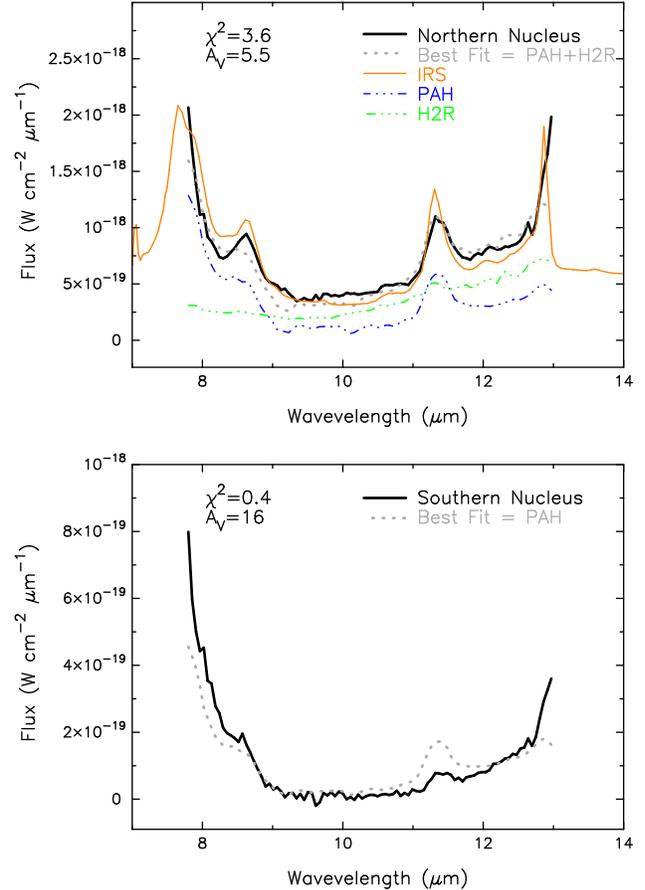

\centering
\includegraphics[angle=270,scale=0.35]{nucleo_norte.cte_Av_fit.ps}\\[8pt]
\includegraphics[angle=270,scale=0.35]{nucleo_sur.cte_Av_fit.ps}
\caption{N-band spectra of the nuclei in NGC\,3256. Absolute flux
calibration was obtained using N-band photometry measured in
1.3\arcsec\ diameter apertures (see Table 1). The Spitzer IRS spectrum
obtained by Brandl et al.~(2006) is overplotted on the TReCS
observations for the northern nucleus, after applying a scaling factor
of 0.18.}
\end{figure}

The N-band spectra of the northern and southern nuclei in NGC\,3256
are shown in Figure 4. Both spectra are flanked by the rising PAH
features at 7.7 $\mu$m and the narrow [NeII]$\lambda12.8\mu$m emission
line which is blended with the 12.7 $\mu$m PAH feature. Unfortunately
the [NeII]$\lambda12.8\mu$m line is not completely covered in the
spectral range of our data but it can be clearly seen in the data of
Mart\'{\i}n-Hern\'andez et al.~(2006).

The nuclear region of NGC\,3256 was observed with the Spitzer IRS
spectrograph. As can be seen in Figure 4, the IRS spectrum is a very
good match to the TReCS observations of the northern nucleus, which
clearly dominates the emission in the mid-IR (see Figure 1). Since the
flux calibration of the IRS spectrum was obtained from a much larger
aperture (10.5\arcsec), a scaling factor of 0.18 had to be applied to
these data. This is also similar to the ratio between the 1.3\arcsec
and 9.5\arcsec\ fluxes measured in the apertures applied to the TIMMI2
images, and presented in Table 1.

Following the model proposed by other authors (e.g., Laurent et
al., 2000; Sturm et al., 2000), we assume a simple description of the
N-band emission, and fitted the spectra of both nuclei as a combination of
several components:

\[ F_{\lambda}= \lbrace C_1F^{HIIR}_{\lambda}+C_2F^{PAH}_{\lambda}+C_3F^{AGN}_{\lambda}\rbrace \times 10^{-A_{\lambda}/2.5} \]

where $F^{HII}$ and $F^{PAH}$ are empirical templates of an HII region
and a PAH-dominated PDR region (Peeters, Spoon \& Tielens, 2004). The
$F^{AGN}$ term is a template of a type 2 AGN, i.e., it corresponds to
the reprocessed emission from the dusty torus without the directly
visible contribution from an active nucleus. These templates were
taken from Sturm et al.~(2000) and Le Floc'h et al.~(2001).
$A_{\lambda}$ is the assumed extinction curve. We have used the mid-IR
extinction curves published by Draine (2003). $C_1 , C_2$, and $C_3$
are free parameters which are determined by using $\chi^{2}$
minimization and the $\sigma$ spectrum, which is dominated by
background emission.
   
\begin{table}
\centering
\caption{Results from the spectral fitting. The percentage contributions of each component
to the integrated 8-12.5 $\mu$m flux are presented.}
\begin{tabular}{cccrrr}
\hline
Nucleus & $\chi^{2}$ & Fixed A$_{\rm V}$ & PAH         & HIIR        & AGN  \\\hline
North   & 3.6        &    5.5      & $\sim 55\%$ & $\sim 45\%$ & 0\%  \\
South   & 0.4        &    16.0     & 100\%       & 0\%         & 0\%  \\\hline
\end{tabular}
\end{table}

Since the spectral range of our data is too narrow to allow us to
determine the strength of the continuum, the PAH features and the
extinction (which includes the silicate absorption at $\sim 9.8\mu$m),
we have instead assumed a fixed value for the normalization of the
extinction curve, ($A_V$). Since our spectra were obtained using a
$\sim 1\arcsec$ slit, we have adopted $A_V = 5.5$ for the northern
nucleus and $A_V = 16$ for the southern nucleus, as was determined
from the 0.3\arcsec resolution NICMOS images (Lira et al.,
2002). These values are rather uncertain, since they were determined
by assuming an intrinsic $H-K$ color for the starburst region. The
same method gives $A_V = 2.5$ and $A_V = 5.3$ magnitudes for the
northern and southern nuclei respectively, when the colours are
measured from a 3\arcsec\ aperture, while the use of a correlation
between the radio 6~cm and [FeII] fluxes implies $A_V = 7.8$ and $A_V
= 10.7$ for the northern and southern nuclei respectively (Kotilainen
et al., 1996).

The outcome from the spectral fitting is summarized in Table 2. A
clear result is that the northern nucleus shows a strong continuum
contribution, while the southern nucleus is dominated by PAH emission,
without the need for a significant continuum component. We have
explored adopting lower values of $A_V$ when fitting the spectrum of
the southern source. In particular, in Section 5 we derive an
extinction of $A_V =10$ for this nucleus. However, these changes have
very little impact on the best fit results.

Our fit suggests that the continuum of the northern nucleus has a
stellar origin ($F^{HII}$), although if only an AGN continuum is
allowed for the fit, then the result is also acceptable ($\chi^{2} \la
4$). However, the presence of an AGN in NGC\,3256 is not supported by
the very considerable body of observational evidence which indicates
that an active nucleus is not present in this galaxy, or at least if
it exists at all it is not the main component responsible for the
powerful IR and X-ray emission (Doyon et al.,1994, Kotilainen etal,
1996; Moorwood \& Oliva 1994, Norris \& Forbes, 1995; Rigopoulou etal,
2000; L\'{\i}pari et al., 2000, Lira et al., 2002, Verma et al.,
2003).

We have also measured the flux and equivalent width of the 11.3 $\mu$m
PAH feature clearly seen in both our spectra. Since it is not possible
to easily unblend this emission from the neighbouring 12.7 $\mu$m PAH
band and strong [NeII]$\lambda12.8\mu$m line, we have derived these
measurements by simply assuming the continuum level to be a straight
line from 10.9 $\mu$m to 11.7 $\mu$m. We compare our measurements with
those from the literature in Table 3. In general there is good
agreement between values obtained through similar slit widths, given
the uncertainties in the absolute flux calibrations for spectroscopy
and the different adopted continuum levels. In particular, we have
re-analysed the IRS data using the continuum fit as defined above.

\begin{table}
\centering
\caption{Flux and EW for the 11.3 $\mu$m PAH emission band.}
\begin{tabular}{lcccc} \hline
Instrument & Slit       & Flux                        & EW       & Ref.\\ 
           & (\arcsec)  & ($\times10^{-20}$ W/cm$^2$) & ($\mu$m) &\\ \hline 
\multicolumn{5}{l}{Northern nucleus}                             \\ 
TIMMI2     & 1.2        & 25                          & --       &1\\ 
TReCS      & 1.3        & 17                          & 0.3      &2\\
TIMMI2     & 3.0        & 30                          & 0.4      &3\\
IRS        & 10.5       & 85                          & 0.5      &4\\
IRS        & 10.5       & 120                         & 0.4      &2\\
ISO        & 24.0       & 107                         & 1.0      &3\\ \hline
\multicolumn{5}{l}{Southern nucleus}                             \\ 
TIIMI2     & 1.2        & 5                           & --       &1\\ 
TRcES      & 1.3        & 2                           & 0.7      &2\\ \hline
\multicolumn{5}{l}{Refs: 1 - Mart\'{\i}n-Hern\'andez et al., 2006; 2 - this work; }\\
\multicolumn{5}{l}{3 - Siebenmorgen et al., 2004; 4 - Brandl et al., 2006.}\\
\end{tabular}
\end{table}

\section{Modelling the Spectral Energy Distribution (SED)}

\begin{figure}
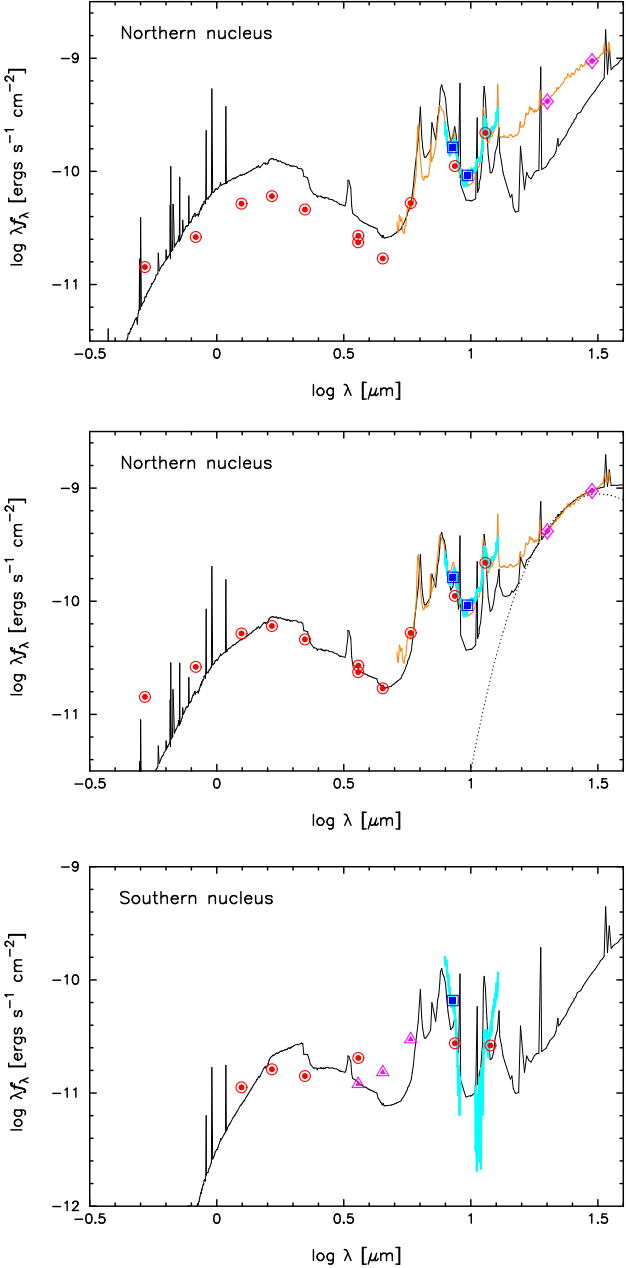

\centering
\includegraphics[angle=270,scale=0.38]{N.dopita_sed.ps}\\[10pt]
\includegraphics[angle=270,scale=0.38]{N.dopita_GB_sed.ps}\\[10pt]
\includegraphics[angle=270,scale=0.38]{S.dopita_sed.ps}
\caption{SED for the nuclei in NGC\,3256. The top and middle panels
show a fit to the northern nucleus (including a Gray Body component in
the middle panel -- dotted line), while the lower panel shows a fit to
the southern nucleus. Optical, NIR and MIR aperture photometry is
shown by circles (see Table 1). The mid-IR TReCS (and for the northern
nucleus, IRS) spectra, scaled to the observed $3\arcsec$ diameter
aperture TIMMI2 fluxes, are also shown with a thick and thin line,
respectively. The spectra were binned to add further points to the
SEDs (squares for TReCS data; diamonds for IRS data). Starburst models
were taken from the work of Dopita et al.~(2005).}
\end{figure}

In Figure 5 we show the SEDs for the nuclei in NGC\,3256. Optical
WFPC2 photometry is presented for the northern nucleus only. For both
nuclei we have included near-IR fluxes from Kotilainen et al.~(1996),
the IRAC photometry, narrow-band Gemini-TReCS observations at 8.74
$\mu$m from Alonso-Herrero et al.~(2006), and our new TIMMI2
observations.

For construction of the SEDs we aimed to use fluxes obtained from a
spatial region closely corresponding to the $3\arcsec$ diameter
apertures used for our TIMMI2 imaging, which is characterised by a
diffraction limited FWHM $\sim 0.8\arcsec$. This is already the case
for the NIR data taken from Kotilainen et al.\ (FWHM $\sim
0.8-1\arcsec$). The WFPC2 and IRAC photometry presented in Section 3
also correspond to effective $3\arcsec$ diameter apertures. Similar to
the analysis performed to determine the IRAC photometry, we estimate
that a factor 1.4 is necessary to correct the photometry presented in
Alonso-Herrero et al.\ using a 1.4\arcsec\ diameter aperture (with a
resolution of $\sim 0.3\arcsec$) to $3\arcsec$. All results from the
aperture photometry are presented in Table 1.

The TReCS spectra are overplotted on the SEDs in Figure 5 after
scaling them to the TIMMI2 flux observed in a 3\arcsec\ aperture (see
Table 1). The IRS spectrum was included only for the northern nucleus
after applying a scaling factor of 0.44. Given the good match between
the TReCS and IRS spectra in the $8-12.5 \mu$m wavelength range (see
Figure 4), we have assumed that the northern nucleus is also
responsible for most of the IRS flux at wavelengths longer than
$13\mu$m. As discussed before, all photometric measurements were
corrected to correspond to fluxes predicted within a $\sim 3\arcsec$
aperture.

Kotilainen et al.~(1996) showed that the southern nucleus becomes
relatively more prominent at longer wavelengths, and is comparable to
the northern nucleus in the L-band. They interpreted this difference
as being due to different amounts of dust obscuration towards the two
nuclei.  However our N-band data show a very weak southern nucleus
when compared with the northern source, which is not consistent with
such an explanation. In addition the IRAC observations presented in
Figure 3 confirm this trend.  The measured fluxes clearly show that
their emission is very nearly the same at $\sim 4.5 \mu$m, but diverge
longwards and shortwards of this wavelength (see Table 1).

We performed an unweighted fit to the SEDs using the models presented
by Dopita et al.~(2005), which include a self-consistent treatment of
dust re-emission from a starburst region. We also added extinction
from a foreground screen of dust which was parameterised using the
Calzetti et al.~(2000) extinction law, ranging from the optical to
$2.2\mu$m, and the Draine et al (2003) extinction curves from
$2.2\mu$m longwards. The best fit models imply visual band extinctions
of $\approx 5$ and $\approx 10$ mags for the northern and southern
nuclei, respectively. This is in good agreement with the previous
determinations discussed in Section 4 within the uncertainties. These
are mainly due to the assumption that a screen of foreground dust can
account for the extinction towards the nuclei all the way from the
optical to the mid-IR, since the geometry of these regions is probably
more complex. Also, the adoption of a particular extinction curve
would have some impact on the final determined extinction, while the
templates used in our description of the sectra might not properly
describe the intrinsic emission.

The starburst models have two free parameters, the ISM pressure (with
$P/k=10^4, 10^6$ and $10^7$ cm$^{-3}$ K) and the characteristic time
scale for the 'clearing' of the starforming regions (i.e., the time
required for the covering factor of the star-forming clouds to
decrease from 0.9 to 0.1). Changes in the pressure affect the model
SEDs for fluxes above $\sim 15\mu$m and therefore are not well
constrained by our SED data. However, Rigopoulou et al.~(1996) have
determined an electron density $\sim 300-400$ cm$^{-3}$ for the
central region in NGC\,3256, and therefore we considered those SED
models corresponding to $P/k=10^{6-7}$ cm$^{-3}$ K. The best fit
models imply clearing times between $16-32$ Myr for both nuclei.

Generally the starburst models provide a good match to the SEDs. The
northern nucleus, however, clearly requires an extra component in its
continuum at $\ga 10\mu$m, as can be seen by comparing the top and
middle panels in Figure 5. This is in agreement with our findings from
the spectral analysis. The reprocessing of emission absorbed by
foreground dust is not expected to contribute significantly at these
wavelengths as the characteristic temperature of this emission would
be lower than 50 K. Instead we have modelled the excess flux as a Gray
body assuming an emissivity with index $\beta = -1.5$. The best fit
temperature for this component is $\sim 80$ K. The origin of this
emission could be explained if the actual distribution of dust grain
sizes is different to the one used in Dopita's models. If this
interpretation is correct then the excess emission is due to a
comparatively larger contribution from very small dust grains, which
dominate the dust re-emission in the mid-IR (D\'esert, Boulanger \&
Puget, 1990).

\section{Discussion}

The normalization of the starburst SED models allow us to infer a star
formation rate (SFR) of $\sim 15$ and $\sim 6 M_{\odot}/$yr for the
northern and southern nuclei, respectively. This estimation does not
take into account the Gray-body component added to the fit of the
northern nucleus since its integrated luminosity corresponds to $\la
25\%$ of the total emission. These SFRs are in good agreement with the
values obtained using the supernova rate of 0.3 yr$^{-1}$ per nucleus
derived from radio observations (Norris \& Forbes 1995) by assuming a
conversion ${\rm SNR} = 0.02 \times {\rm SFR}$ as inferred from the
Starburst\,99 models using a constant star formation episode
(Leitherer et al., 1999).

One of the interesting results obtained from our new N-band data is
establishing the photometric and spectroscopic differences between the
northern and southern nuclei, which cannot be explained simply by
differences in dust obscuration. Since the starburst activity in these
nuclei must have been triggered by the merger of two progenitors, the
age of the starbursts is likely to be comparable, but other physical
parameters could be different in the two separate merging parent
galaxies.

X-ray, optical and near-IR observations are indicative of a highly
obscured southern nucleus which is completely hidden at visible
wavelengths but which becomes increasingly prominent in the near-IR.
The brightness of the two peaks are comparable at $\sim 4 \mu$m and at
radio wavelengths. Hence, it might be anticipated that in the mid-IR,
where extinction becomes negligible, the two nuclei would also have
similar flux levels. But this is not observed. Instead, we find that
the northern nucleus exhibits excess continuum emission at wavelengths
$\ga 10\mu$m when compared with the southern nucleus and with our
model SEDs.

Interestingly a similar behaviour was found by Alonso-Herrero et
al.~(2006) when comparing the $8\mu$m and Pa$\alpha$ emission from a
sample of LIRG nuclei, including NGC\,3256. While the northern nucleus
shows a high $8\mu$m/Pa$\alpha$ ratio, as was also seen in other
LIRGs, the lower $8\mu$m/Pa$\alpha$ value of the southern nucleus is
consistent with the ratios observed in HII regions. Alonso-Herrero et
al.~(2006) explain this difference by the presence of large scale
diffuse emission, which is characterised by strong PAH features in the
mid-IR. As can be seen in Table 3, for NGC\,3256 the EW of the 11.3
$\mu$m PAH feature changes at most by a factor 3 between small
(11.2--1.3\arcsec) and large (10.5--24\arcsec) apertures, while the
line fluxes are greater by a factor $\sim 6$. This suggests that the
fractional increase in the continuum flux is also substantial (factor
$\ga 2$). In addition, the data presented in Figures 4 and 5 suggest
that a significant fraction of the excess mid-IR flux is due to
continuum emission produced by warm dust, heated in-situ by the
starburst itself.

Using Spitzer IRS observations, Brandl et al.~(2006) has quantified
the continuum fluxes at 6, 15 and 30 $\mu$m for a sample of starburst
galaxies, including NGC\,3256. Brandl's observations show that the
flux ratio $F_{6\mu{\rm m}}/F_{15\mu{\rm m}}=0.14$ determined for
NGC\,3256 is significantly lower than the average $<F_{6\mu{\rm
m}}/F_{15\mu{\rm m}}>=0.26$ found for their starburst sample. As we
have shown, the northern nucleus is more luminous that the southern
nucleus by a factor of $\sim 2$ at 5.8 $\mu$m, while at 15$\mu$m the
flux should be dominated by the northern nucleus. Indeed, using the
best fit SEDs we find $F_{6\mu{\rm m}}/F_{15\mu{\rm m}}=1.2$ for the
southern nucleus, while for the northern nucleus we adopt the value
tabulated from the IRS spectrum. The full range of observed flux
ratios in Brandl's sample is $\sim 0.1-0.6$, which probably reflects
the range in starburst conditions, such as age and dust properties. We
find that the nuclei in NGC\,3256 are representative of the two
extremes of this distribution, but are found in the same central
region of a single galaxy.

Finally we compare the 10.5$\mu$m and (unabsorbed, 2-10 keV) X-ray
fluxes for both nuclei with the correlation derived by Krabbe, B\"oker
\& Maiolino (2001) using N-band MANIAC observations. The X-ray fluxes
correspond to $9 \times 10^{-17}$ and $4 \times 10^{-17}$ W m$^{-2}$
(Lira et al., 2002), while the $10.5\mu$m fluxes ($\Delta \lambda = 5
\mu$m, assuming the best fit SEDs) are $\sim 400$ and $\sim 40$ mJy,
respectively. It is found that both nuclei are X-ray deficient given
their N-band photometry. In particular, because of its high mid-IR
flux, the northern nucleus becomes a clear outlier in this
correlation. But as shown by Krabbe's work, this is not the case for
NGC\,3256 when the fluxes are integrated over a large region, implying
that the correlation holds for global starburst properties, but
individual regions can show a much wider range of properties. In
particular, our work shows that in-situ dust heating in LIRGs, whose
emission can dominate the mid-IR emission, can be confined to very
compact (few hundred-pc scale) regions.

\section*{Acknowledgements}

P.L.~and V.G.~are grateful of support by the Fondecyt project No
1040719. We would like to thank Dr.~B.~Brandl for kindly supplying the
IRS observations. We also wish to thank an anomymous referee for
useful comments.

{}

\end{document}